\documentstyle[amsfonts,12pt]{article}
\begin{document}

\title{ Coherent-state path-integral approach for constrained fermion
  systems}

\author{Georg Junker\\
Institut f\"ur Theoretische Physik, Universit\"at
  Erlangen-N\"urnberg,\\ Staudtstr.\ 7, D-91058 Erlangen, Germany\\
E-mail: junker@theorie1.physik.uni-erlangen.de\\
~\\
John R.\ Klauder\\
Departments of Physics and Mathematics, University of Florida,\\
  Gainesville, FL-32611, USA\\ E-mail: klauder@phys.ufl.edu}  

\maketitle

\begin{abstract}
The coherent-state path-integral representation for the propagator of
fermionic systems subjected to first-class constraints is constructed. 
As in the bosonic case the importance of path-integral measures for Lagrange
multipliers is emphasized. One example is discussed in some detail.
\end{abstract}
%%%%%%%%%%%%%%%%%%%%%%%%%%%%%%%%%%%%%%%%%%%%%%%%%%%%%%%%%%%%%%%%%%%%%%%%%%%%
\section{Introduction}
The quantization of constrained systems is of considerable importance in many
areas of theoretical physics. At the last meeting of this conference series,
held at Dubna (Russia) in 1996, one of us has reexamined this problem from
the point of view of coherent-state path integrals.\cite{K96,K97} It is the
aim of this contribution to show that this work on bosonic systems can be
generalized to systems with fermionic degrees of freedom. Here we will limit
ourselves to a brief discussion of first-class constraints. More details,
including also an extensive discussion of second-class constraints, have
already been published elsewhere.\cite{JK98} Similar work has also been
studied by Prokhorov and Shabanov.\cite{PS97} 

The fermionic quantum systems considered here consist of $N$ fermionic degrees
of freedom characterized by annihilation and creation operators obeying the
anticommutation relations $\{f_i,f_j\}=0$, $\{f^\dagger_i, f_j^\dagger \}=0$,
$\{f_i^\dagger , f_j\}=\delta_{ij}$ and acting on the ``$N$-fermion'' Hilbert
space ${\cal H}\equiv{\mathbb C}^{2^N}={\mathbb
  C}^2\otimes\cdots\otimes{\mathbb C}^2$. The quantum dynamics of such a
system is completely characterized by a 
normal-ordered self-adjoint Hamiltonian $H=H(f^\dagger,f)$ and the associated
constraints, which are denoted by $\Phi_a(f^\dagger,f)$ and
$\chi_\alpha(f^\dagger,f)$ for constraints being even and odd in the fermion
operators, respectively. Here $f$ and $f^\dagger$ stand for the ordered set
$\{f_1,\ldots,f_N\}$ and $\{f^\dagger_1,\ldots,f^\dagger_N\}$,
whereas $a$ and $\alpha$ enumerate the even and odd constraints,
respectively. As for the Hamiltonian, we 
assume that the operator-valued constraints are self-adjoint and normal
ordered. The associated normalized fermionic coherent states
$|\Psi\rangle\equiv 
|\psi_1 \cdots \psi_N\rangle=|\psi_1\rangle\otimes\cdots\otimes|\psi_N\rangle$
are labeled by odd Grassmann variables, satisfying \ $\psi_i^2=0=\bar{\psi}_i^2$, and
fulfill the relation $\langle\Psi''|O(f^\dagger,f)|\Psi'\rangle
=O(\bar{\Psi}'',\Psi')\langle\Psi''|\Psi'\rangle$ for any normal-ordered
operator $O$. We also note that these states form an over-complete set,
$\langle\Psi''|\Psi'\rangle=\exp\{-\frac{1}{2}\bar{\Psi}''\cdot(\Psi''-\Psi')
+\frac{1}{2}(\bar{\Psi}''-\bar{\Psi}')\cdot\Psi'\}$, where
$\bar{\Psi}''\cdot\Psi'\equiv\bar{\psi}''_1\psi'_1+\cdots+
\bar{\psi}''_N\psi'_N$ 
etc., and admit a resolution of unity in ${\cal H}$, 
$\int d\bar{\Psi}d\Psi \,|\Psi\rangle\langle\Psi|=1$, based on the standard
definition of Grassmann integration.\cite{JK98,KS85} 
%%%%%%%%%%%%%%%%%%%%%%%%%%%%%%%%%%%%%%%%%%%%%%%%%%%%%%%%%%%%%%%%%%%%%%%%%%%%
\section{First-class Constraints}
In this contribution we limit ourselves to the case of first-class
constraints, i.e.\ the constraints together with the Hamiltonian are assumed
to close a Lie superalgebra ($c,d,g,h$ and $k$ denote complex-valued
structure constants) 
\begin{eqnarray}
  \label{algebra1}
  &&[\Phi_a,\Phi_b]=i\,c_{ab}{}^c\Phi_c\;,\quad
  [\Phi_a,\chi_\alpha]=i\,d_{a\alpha}{}^\beta\chi_\beta\;,\quad
  \{\chi_\alpha,\chi_\beta\}=i\,g_{\alpha\beta}{}^a \Phi_a\;,\\
  \label{algebra2}
  &&[\Phi_a,H]=i\,h_{a}{}^b\Phi_b\;,\quad
  [\chi_\alpha,H]=i\,k_{\alpha}{}^\beta\chi_\beta\;.
\end{eqnarray}
The physical subspace of the Hilbert space ${\cal H}$ is defined by
\begin{equation}
  \label{Hphys}
  {\cal H}_{\rm phys}=\{|\varphi\rangle\in{\cal H}\,:\,
  \Phi_a|\varphi\rangle=0\mbox{ and }\chi_\alpha |\varphi\rangle=0 \mbox{ for
  all } a \mbox{ and } \alpha\} \;.
\end{equation}
However, from the last relation in (\ref{algebra1}) it is obvious that the
even constraints, $\Phi_a|\varphi\rangle=0$ for all $a$, imply the odd
constraints $\chi_\alpha |\varphi\rangle=0$ and, therefore, it suffices to
consider only the Lie algebra, c.f.\ the first relation in (\ref{algebra1}),
of the even constraints generating the Lie group $G$. As a consequence, this
subset of (boselike) constraints may now be treated in analogy to the
bosonic case.\cite{K96,K97} 
That is, we introduce the projection operator ($\mu$ denotes the invariant
normalized Haar measure on $G$)
\begin{equation}
  \label{projector}
  {\mathbb E}\equiv\int_G d\mu(\xi)\,\exp\{-i\xi^a\Phi_a\}={\mathbb
  E}^\dagger={\mathbb E}^2 =\exp\{-i\eta^a\Phi_a\}{\mathbb E}\;,
\end{equation}
where the last relation is valid for any values of the real group parameters
$\eta$. Clearly, the above definition only holds for compact groups
$G$. However, even in the case of non-compact groups we may strictly follow
the methods designed in the bose case.\cite{K97}
Obviously, we have ${\cal H}_{\rm phys}={\mathbb E}{\cal H}{\mathbb E}$ and
the constrained time-evolution operator admits the following representations
(by using the first relation in (\ref{algebra2}) and the last one in
(\ref{projector}): 
\begin{equation}
  \label{timeevol}
  \exp\{-itH\}{\mathbb E}={\mathbb E}\exp\{-itH\}={\mathbb
  E}\exp\{-itH\}{\mathbb E}={\mathbb E}\exp\{-it({\mathbb E}H{\mathbb
  E})\}{\mathbb E}\;.
\end{equation}
For the fermion coherent-state matrix element of this operator, that is, the
propagator $\langle\Psi''|\exp\{-itH\}{\mathbb E}|\Psi'\rangle$, we will now
construct a path-integral representation again following closely the previous
approach.\cite{K97}  
%%%%%%%%%%%%%%%%%%%%%%%%%%%%%%%%%%%%%%%%%%%%%%%%%%%%%%%%%%%%%%%%%%%%%%%%%%%%
\section{Path-Integral Representation of the Constrained Propagator}
In order to construct a path-integral representation for the constrained
propagator we start with the last relation in (\ref{projector}), divide the
time interval $t$ into $M$ short-time intervals $\varepsilon=t/M$ and then
use the group composition law following from the first relation in
(\ref{algebra2}):
\begin{equation}
  \label{path1}
  e^{-itH}{\mathbb E} =
  e^{-itH}e^{-i\xi^a\Phi_a}{\mathbb E} =
 \prod_{n=1}^{M}e^{-i\varepsilon H}
                e^{-i\varepsilon\eta^a\Phi_a}{\mathbb E}\;.
\end{equation}
A time-lattice path-integral representation of the corresponding propagator
then immediately follows from an $M$-fold insertion of the resolution of unity
with fermion coherent states and the limit $\varepsilon\to 0$ such that
$M\varepsilon=t$ remains fixed ($\Psi_M\equiv\Psi''$):
\begin{eqnarray}
&&\langle\Psi''|\exp\{-itH\}{\mathbb E}|\Psi'\rangle = \lim_{\varepsilon\to 0}
  \prod_{n=0}^{M-1}\int d\bar{\Psi}_n d\Psi_n
  \int_G d\mu(\xi)\,\langle\Psi_0|e^{-i\xi^a\Phi_a(f^\dagger,f)}|\Psi'\rangle
  \nonumber \\
&&\hspace{20mm}\times
  \prod_{n=1}^{M}\exp\left\{\textstyle
    -\frac{1}{2}\bar{\Psi}_n\cdot(\Psi_n-\Psi_{n-1})+
    \frac{1}{2}(\bar{\Psi}_n-\bar{\Psi}_{n-1})\cdot\Psi_{n-1}\right. 
    \nonumber\\
&&\left.\hspace{37mm}
   -i\varepsilon H(\bar{\Psi}_n,\Psi_{n-1})\rule{0mm}{3mm}
   -i\varepsilon\eta_n^a\Phi_a(\bar{\Psi}_n,\Psi_{n-1})\right\}\;.\label{path2}
  \end{eqnarray}
Here we note that despite the fact that the time-dependent Lagrange
multipliers $\eta^a$ explicitly appear on the right-hand side the final result
clearly does not depend on them. Hence, we are free to average this expression
by any, in general complex-valued, normalized measure 
$\prod_n \int dC(\eta_n)=1$. Therefore, we may express the above path integral
formally by
\begin{equation}
 \int{\cal D}\bar{\Psi}{\cal D}\Psi{\cal D}C(\eta)
  \exp\left\{i\int_0^td\tau\textstyle
  \left[\frac{i}{2}(\bar{\Psi}\cdot\dot{\Psi}-\dot{\bar{\Psi}}\cdot\Psi)
        -H(\bar{\Psi},\Psi)-\eta^a\Phi_a(\bar{\Psi},\Psi)\right]\right\}
\end{equation}
with the only requirement that the measure ${\cal D}C(\eta)$ should at least
introduce one projection operator in order to respect the constraints.
%%%%%%%%%%%%%%%%%%%%%%%%%%%%%%%%%%%%%%%%%%%%%%%%%%%%%%%%%%%%%%%%%%%%%%%%%%%%
\section{Example}
As an elementary but instructive example we choose two fermionic degrees of
freedom $N=2$ subjected to the constraint $\Phi(f^\dagger,f)=f_1^\dagger
f_2+f_2^\dagger f_1$ and a vanishing Hamiltonian $H=0$. In this case the
projector (\ref{projector}) has the following representations
\begin{equation}
  {\mathbb E}=\frac{1}{2\pi}\int_0^{2\pi}d\xi\,e^{-i\xi\Phi}=1-\Phi^2
        =1-f_1^\dagger f_1- f_2^\dagger f_2+2f_1^\dagger f_1f_2^\dagger f_2
\end{equation}
and its coherent-state matrix element $\langle\Psi''|{\mathbb E}|\Psi'\rangle$
is represented by the following path-integral
($\Delta\Psi_n=\Psi_n-\Psi_{n-1}$,
$\Delta\bar{\Psi}_n=\bar{\Psi}_n-\bar{\Psi}_{n-1} $)
\begin{equation}
  \begin{array}{l}
\displaystyle
 \int{\cal D}\bar{\Psi}{\cal D}\Psi{\cal D}C(\eta)
  \exp\left\{i\int_0^td\tau\textstyle
  \left[\frac{i}{2}(\bar{\Psi}\cdot\dot{\Psi}-\dot{\bar{\Psi}}\cdot\Psi)
        -\eta(\bar{\psi}_1\psi_2 +\bar{\psi}_2\psi_1 )\right]\right\}\\
\displaystyle
\mbox{  }=\lim_{\varepsilon\to 0}\prod_{n=0}^{M}\int d\bar{\Psi}_n d\Psi_n
          \prod_{n=1}^{M}\int_{-\infty}^{\infty} d\eta_n\, \delta(\eta_n)
          \int_0^{2\pi}\frac{d\xi}{2\pi}\,
              \langle\Psi_0|e^{-i\xi\Phi(f^\dagger,f)}|\Psi'\rangle\\
\displaystyle
\mbox{~~~~~~~~~}\times\prod_{n=1}^{M}\exp\left\{\textstyle 
                       -\frac{1}{2}\bar{\Psi}_n\cdot\Delta\Psi_n
                       +\frac{1}{2}\Delta\bar{\Psi}_n\cdot\Psi_{n-1}
                   -i\varepsilon\eta_n\Phi(\bar{\Psi}_n,\Psi_{n-1})\right\}\;.
  \end{array}
\end{equation}
Our choice for the measure of the Lagrange multipliers is obvious and
obeys the requirement to insert at least one projection operator (here at the
initial stage). Integration over the $\eta$'s is trivial and for that over the
$\Psi$'s we use the convolution formula
\begin{equation}
  \begin{array}{l}
\displaystyle
  \int d\bar{\Psi}_n d\Psi_n\; e^{
  -\frac{1}{2}\bar{\Psi}_n\cdot\Delta\Psi_n
  +\frac{1}{2}\Delta\bar{\Psi}_n\cdot\Psi_{n-1}}
 \; e^{ -\frac{1}{2}\bar{\Psi}_{n+1}\cdot\Delta\Psi_{n+1} 
      +\frac{1}{2}\Delta\bar{\Psi}_{n+1}\cdot\Psi_{n}}\\
\mbox{~~~~~~~}=\exp\left\{\textstyle
      -\frac{1}{2}\bar{\Psi}_{n+1}\cdot(\Psi_{n+1}-\Psi_{n-1})+
      \frac{1}{2}(\bar{\Psi}_{n+1}-\bar{\Psi}_{n-1})\cdot\Psi_{n-1}\right\}\;.
  \end{array}
\end{equation}
Finally, we note that
($|\Psi_0\rangle\equiv|\psi_1\rangle\otimes|\psi_2\rangle$) 
\begin{equation}
  \begin{array}{l}
\langle\Psi_0|e^{-i\xi(f_1^\dagger f_2 +f_2^\dagger f_1)}|\Psi'\rangle=
\langle\Psi_0|\Psi'\rangle\\[1mm]
\mbox{~~~}\times\left[
1+(\cos\xi-1)(\bar{\Psi}_0\cdot\Psi'+2\bar{\psi}_1\bar{\psi}_2\psi_1'\psi_2')
-i\sin\xi\,(\bar{\psi}_1\psi_2'+\bar{\psi}_2\psi_1')\right]\;,
  \end{array}
\end{equation}
which immediately gives rise to the explicit result
\begin{equation}
  \begin{array}{ll}
  \langle\Psi''|{\mathbb E}|\Psi'\rangle & =
e^{-\frac{1}{2}\bar{\Psi}''\cdot\Psi''}
e^{-\frac{1}{2}\bar{\Psi}'\cdot\Psi'}
\left[1-3\bar{\psi}''_1\bar{\psi}''_2\psi_1'\psi_2'\right]
\\[1mm]
& = \langle\Psi''|\Psi'\rangle \left[1-\bar{\Psi}''\cdot\Psi'-
                2\bar{\psi}''_1\bar{\psi}''_2\psi_1'\psi_2'\right]\;.
  \end{array}
\end{equation}
Further examples of first-class as well as second-class constraints can be
found elsewhere.\cite{JK98}

\end{document}